> "... we want more than just a formula. First we have an observation, then we have numbers that we measure, then we have a law which summarizes all the numbers. But the real *glory* of science is that *we can find a way of thinking* such that the law is *evident*."
>
> "The Feynman lectures on physics",
> Addison-Wesley, MA, 1966, p.26-3.

# THE GHOSTLY SOLUTION OF THE QUANTUM PARADOXES AND ITS EXPERIMENTAL VERIFICATION[*]

Raoul Nakhmanson

Frankfurt am Main, Germany[†]

This conference is entitled "Frontiers of fundamental physics". What does this mean? Is it the frontiers of today's physical knowledge, or is it the frontiers of physics itself as a science?

In my paper I shall try to show that today it is the same: the frontiers of contemporary physical knowledge coincide with the conceptual frontiers of physics as a science regarding the behaviour of so-called inanimate matter and even cross over to invade into the kingdom of ghost. Such a point of view permits a very natural interpretation of quantum phenomena, and suggests essentially new experiments in which information plays the principal rôle.

The microworld has surprised the "classical" physicists with the following paradoxes:[1,2]

1) Before quantum mechanics (QM) was created: quantization of mass, charge, energy, angular momentum; the identity of particles of the same type; wave-particle duality.

2) In QM: statistical predictions, Heisenberg's uncertainty principle, Pauli's exclusion principle.

3) In standard (Copenhagen) interpretation of QM: rejection of the classical realism, a ban on speaking about non-measured parameters, trajectories, etc.; Bohr's complementarity principle, collapse of the wave function.

The Copenhagen interpretation is only a translation of the mathematical formalism of QM to the ordinary language but not an interpretation in a common sense, because it does not explain how, why, and in which frames this formalism works. Feynman told his students that the quantum world was not like anything that we know; and although everybody knows QM, many people use it, some of them develop it, but nobody understands it.

In discussions about QM the "Gedankenexperimente" play an important rôle. We will discuss three of them which were really performed:

1) Delayed-choice experiment.[3] In one arm of an interferometer a Pockels cell is placed which closes the path of photons at the short moment when they can pass the cell. In accordance with old local-realistic concept each photon flies only in one arm of the interferometer. If it is the arm with the cell the photon will be absorbed and nothing will be registered. If it is another arm, the short work of the cell placed far away does not act on the photon and the same interference as without the cell must be registered. But no interference was found in accordance with QM.

---

2) Aharonov-Bohm effect.[4] In accordance with QM the frequency of wave-function oscillation depends on the energy. If the particle has different energies in different arms of the interferometer, it leads to an additional phase shift and changes the interference pattern. The experiments were performed with an electron interferometer and a magnetic vector potential and justified the predictions of QM. It is of interest that in the experiments the electrons did not cross the magnetic field. From the old classical point of view it looks like non-local action at a distance.

3) Einstein-Podolsky-Rosen (EPR) experiment. It was suggested in[5] and modernized by Bohm.[6] Here two particles emitted simultaneously have common non-factorisable wave function and are measured after parting by a large distance. There is some correlation between the results measured. Bell has shown[7] that any local realistic theory (i.e. theory with hidden parameters and restricted velocity of interaction) estimates the uppermost limit of such correlation, and this limit is smaller than predicted by QM. The experiments being performed[2] are in accordance with QM, and today's dominant opinion is that local realism has been disproved and one must refuse either reality lying beyond the measurements (like Copenhagen) or locality. Later I will show that this conclusion as well as Bell's theorem itself do not have the generality being ascribed to them.

The EPR-scheme raises a question about separability. "Common sense" prompts that after some time and distance the "magic" correlation between particles must disappear, i.e. the factorisation of the wave function must take place. But how? The analogous question is connected with measuring procedure itself: If interaction between particle and apparatus allows several output results, the QM forecasted end state is a superposition of these results. But in practice the result of each measurement is a pure state, and the result of the series is a statistical mixture. It seems as if QM does not describe the whole measurement process.[8]

There are some explanations of the EPR paradox. From the *Copenhagen* point of view it is so as it is. Speaking about some hidden parameters of particles, e.g. directions of spins, before the measurement, has no sense, and Bell's theorem and experiments justify this.

*Non-local theories with hidden parameters*.[9] Here an instantaneous action at a distance is provided by instantaneous collapse of the wave function in all space. The critics emphasize that these theories only rewrite the Schrödinger's equation in a more complex form, giving the same results and nothing new.

*Action of future on the past*.[10] If such action is possible, the future conditions of measurement can act on the hidden parameters of particles at the moment of their departure to tune them for correct correlation. Up to now there is no complete theory ready to defy critique. But common sense prompts that such a world can not be stable.

*Fatalism*. This possibility was noted particularly by Bell.[11] In the spirit of Laplace it is possible to think that everything is pre-determined, particularly our choice of position of analyser. Here we are confronted with the old problem of "free will". If free will exists man (and not only he) can control the choice of alternatives taking into account physical and social conditions. The following chain of syllogisms supports the existence of free will:

      → Useful changes are selected and consolidated by evolution.
      + During evolution the volume of the human brain increases.
      = The volume of brain is a useful quantity.
      + Intelligence depends on the brain volume; as a rule,
         the greater the volume, the higher the intelligence.
      = Intelligence is useful.
      + Intelligence can develop itself only if it can choose among several
         alternatives; only in such situations can intelligence be useful.
      = *Free choice, i.e. free will exists.*

One can reply that the increase of the brain volume as well as evolution itself are included in the fatalistic scenario. But if one considers the existence of free will *ad hoc* as an axiom, then, in accordance with these syllogisms, *free will gives intelligence a chance to evolve.*

The roots of free will do not lie in the macroworld which is ruled by deterministic laws. They lie in the microworld, and quantum uncertainty points to it. Human intelligence is not the only product of free will. It is possible that earler, the free will created some intelligence at the level of its roots, i.e. in microworld. Because the time (measured not in seconds but in events) flowed there much faster, this intelligence had a longer evolution period. Perhaps the golden age of it is over, and now we have to do it only with a "rudimentar" intelligence (so called by Cochran[12]). The additional pointers on intelligent matter are the Einstein's formula $E = mc^2$, the informational character of the wave function $\psi$, the principle of the least action, and quantum-mechanical stochastics.[13]

The development of quantum physics was a step across the boundary between matter and ghost drawn by Descartes. Physicists felt it and spoke about the free will of electrons and ghost (spirit, consciousness, intelligence) in matter. Similar meanings were expressed by Charles Galton Darwin, Eddington, Heisenberg, Schrödinger, Pauli, Jordan, Margenau, Wigner, Charon, Cochran, and others. Feynman said that it looks as if a computer is in each point of space. Cambrige University Press has published a book touching this theme[11] containing interviews with Bell, Bohm, Wheeler, Peierls, Aspect, and others.

Some interesting analogies between microworld and people have been noticed. Niels Bohr saw the manifestation of his complementarity principle in human thinking. Margenau wrote about Pauli's exclusion principle:[14]

> "Prior to that time, all theories had affected the individual nature of so-called 'parts'; the new principle regulated their social behaviour... The particles, though initially assumed to be free, are seen to avoid each other... In a crude manner of speaking, each particle wants to be alone; each runs away then it 'smells' the other, and its sense of smell is keener the more nearly its velocity equals to the other's."

This was said about Fermi-particles. Such behaviour is typical for scientists: each of them tries to find his own theme. Sometimes people's behaviour is like a Bose-particle. Phenomena such as fashion in dress or music, and applause or coughing in concert halls, are examples of Bose-condensation. The same man can manifest himself as a Bose- or Fermi-person. For particles this was only possibe in "big bang" time. Are we now at the same stage of evolution?

The next example concerns the EPR-experiment. Let us suppose there are twins, Ralf and Rolf, both of whom live in Frankfurt and work for Lufthansa as pilots. They fly all over the world but mainly to England and Greece. For Lufthansa (not for their families!) they are indistinguishable "particles". The twins always try to dress alike, they believe that this brings them happiness. Because they are often in different countries, they agree on an order of sartorial priority: cold before warmth and rain before dry spell.

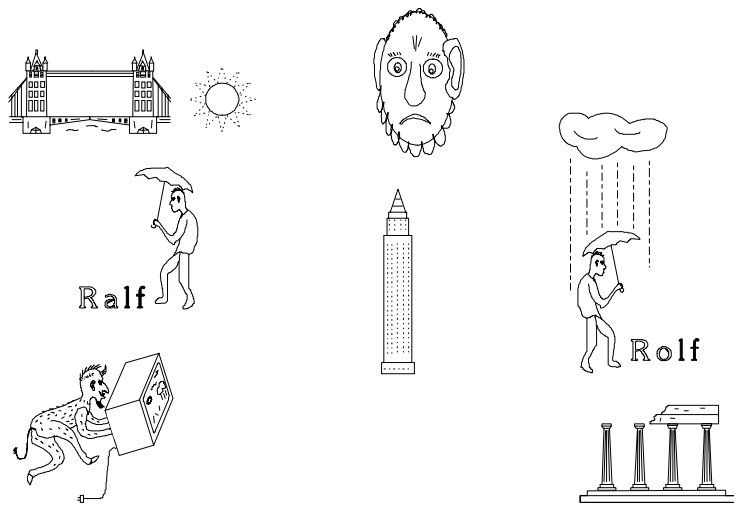

Figure 1. Einstein-Podolsky-Rosen (EPR) experiment and the apparent non-local interaction.

God, who is observing the twins, sees as a rule the striking correlation: the twins dress alike! For example, Ralf and Rolf arrive in England and Greece, respectively. If it is cold in England, not only Ralf but also Rolf wears the overcoat in spite of the warm weather; if it is raining in Greece, not only Rolf but also Ralf hides beneath an umbrella, regardless of whether it is raining or not (Fig.1); etc.. "What is the matter?" - thinks God, - "I estimate experimental conditions, namely, weather in England and Greece and the twins' financial status, telephoning is too expensive for them. It seems there is a non-local interaction between the twins. I am sure it is a new escapade of the devil!"

God's conclusion was only half true. In his heavenly chariot he fell behind the technical progress of the 20th century. He was right suspecting the devil. But up to now the devil does not realize non-local interaction. Instead, he has invented television, power computer for meteorology, and communication satellite. Because of it the twins watch TV at every evening for a good tomorrow world weather *forecast* .

Although our behaviour occurs in real space-time, the strategy of it is not there. It is in our consciousness, which controls our behaviour, taking into account physical and social laws and circumstances. To develop a strategy we use our knowledge only about the past, and propagate it on the future. The thoroughness of the forecast depends on the information taken into account and the power of the intelligence.

But let us come back to physics. Unfortunately the idea about intelligent matter is not developed up to now. They, who spoke about ghost in matter, did not go beyond such a statement and did not suggest any hypotheses and schemes which could be tested experimentally. From another side physicists using QM do not see the necessity of such an idea and follow the principle thought as old as Aristoteles but named after William of Ockham "Ockham's Razor":

"entities should not be multiplied beyond necessity".

Niels Bohr said, that there are trivial and deep statements. To be asked "What is a deep statement?" he answered: "It is such a statement, that an opposite statement is also a deep one." If one accepts the Ockham's principle as a deep statement then, according to Bohr,

"entities should not be canceled beyond necessity"

must also be accepted as a deep statement. Besides, the practical necessity is not the only or main criterion of theory.

A consistent development of the idea of intelligent matter naturally interprets quantum paradoxes as well as QM itself within the limits of local realism, and suggests essentially new experiments with microparticles and atoms in which information plays the principal rôle.

In the new conception the wave function $\psi$ is a strategy-function. It reflects an optimal behaviour of particles. It is not in the real 3-dimensional space. It is in imaginary configuration space, which, in its turn, is in the imagination (consciousness) of the particle. When the particle receives new information (it takes place by any interaction with micro- or macro-objects), it can change its strategy. Thus occurs the collapse of the wave function. It occurs not in the real (infinite) space, but in the consciousness of particles. The consequent time is determined by the rapidity of this consciousness. Therefore, compared with space-time conditions of experiment, collapse is local and instantaneous. Von Neumann and Wigner suggested that human consciousness has influence on the collapse of $\psi$- function. It is not so: in the human consciousness only the human knowledge about the $\psi$- function collapses. The laws of both collapses lie beyond physics.

The wave-particle duality is a mind-body one. In the space there exists only the particle; the wave exists in its consciousness, as well as the reflection of the whole world. If there are many particles, their distribution in accordance with the $\psi$- function looks as a real wave in real space.

Particles are artifical things. Division into different sorts or species with internal identities is typical for mass products. It simplifies production, usage, repairs, and replacement of such objects. Technics, plants and animals illustrate it very well. In the last two cases the production is ruled at the genetic level. For example, people have a very narrow statistical distribution of sizes and masses; the world records in sport differ from the middle results not more than twice. The identity of particles of one sort in QM is analogous to the identity of vehicles of one sort with respect to traffic rules. The individual differences lies beyond QM.

Because of free will the behaviour of particles is not strictly determined. In situations allowing alternative outputs the theory gives only a distribution of priorities. Taking this into account the particle makes its choice. The optimal tactics of proportional proving of all possibilities by an ensemble of disconnected particles is randomization of this choice. To do it the particles must have the generators of random signals.

If some theory and random generator (RG) are used to choose the alternative, it looks like a complete algorithm. Well, but where is the free will now? Is it only to change the RG?

The answer is, that purpose and means create a dependence. Really free is he who has no purpose and no desire including the desire of freedom. Therefore there is a danger that in the "Konsumgesellschaft" we transform ourselves into some kind of automata. Perhaps the microworld did not avoid it. But the new turn of development can be connected with a change of purpose or new information. Besides, Gödel's theorem prompts, that the space of correct statements can be manifold. In such a case to reach a new fold one must make a "quantum jump". "Do not sin against logic, one reaches nothing new", - said Einstein.

Pauli's principle and Bose- and Fermi-particles were discussed above. These types of social behaviour are optimal for searching (fermions) and power action (bosons). In the last case some macroscopical effects can be observed (in superconductors, superfluids, lasers).

With respect to Heisenberg's uncertainty principle: In the new conception it reflects not the reality but QM as a theory of measurement. In reality the particle has definite coordinates, impulse, trajectory, etc.. But during an interaction with the measurement apparatus it has a possibility to choose the next state. It solves this problem using its intelligence (reflected in the $\psi$- function), random generator, and freedom (e.g. reflected in the choice of RG). Neither QM nor any other theory predicts a particular result: it would be a refusal of free will.

In spite of this, the dream of Einstein and other realists, to know the values of all parameters included in a theory can become true. Particles remember what happened and tell it to others. To do this, they must have synchronised clocks, measure rules, and reference points for space and time. In this sense it is possible to speak about absolute coordinates and time, like Greenwich's ones. If we can communicate with particles[13,15], they can say everything about their parameters and forecast their and our future.

The new concept includes the previous realistic ones: *empty waves*[2] and *parallel worlds*[16] exist, but not in the real world: as virtual possibilities they exist in the consciousness of a particle. Not the real[10] but a forecasted *future acts on the past*. The above mentioned danger of total algorithmisation looks like a stochastic *fatalism*.

The new explanation of delayed-choice and EPR experiments, and suggestions how to have "non-QM-results", were done in[13]. The essence is, that particles are well informed about the world and its development. The Aharonov-Bohm effect has the same explanation. Besides, this effect emphasizes a priority of potential against field (in classical physics they enjoy equal rights). From the new point of view it is natural, because potential just contributes to the action function whose minimum as a function of trajectory is wanted. It should be observed that the idea of *forecasting* the conditions on this trajectory is also included in the least action principle. The change from integral form to a differential one does not solve the problem: the notion of derivative is connected with two points, and if we are in one of them, we know only the past conditions in the second point and must extend this into the present.

The proof of Bell's theorem is based on the next assertion: if a particle *1* is measured in the point *A* having a condition (e.g. angle of analyser) $\alpha$, and $P_a$ is a probability of result *a*, then a condition $\beta$ existing in a distant point ***B***, there is a measured particle ***2***, has no influence on the $P_a$, and vice versa. Here Bell and others saw the indispensable requirement of local realism. Mathematically it can be written as

$$P_{ab}(\lambda_{1i},\lambda_{2i},\alpha,\beta) = P_a(\lambda_{1i},\alpha) \times P_b(\lambda_{2i},\beta) \quad , \qquad \text{(Bell)} \qquad (1)$$

where $P_{ab}$ is the probability of the join result *ab*, and $\lambda_{1i}$ and $\lambda_{2i}$ are hidden parameters of particles ***1*** and ***2*** in an arbitrary local-realistic theory. Under the influence of Bell's theorem and the following experiments some "realists" reject locality. In this case an instantaneous action at a distance is possible, and one can write

$$P_{ab}(\lambda_{1i},\lambda_{2i},\alpha,\beta) = P_a(\lambda_{1i},\alpha,\beta) \times P_b(\lambda_{2i},\beta,\alpha) \quad . \quad \text{(non-locality)} \qquad (2)$$

In principle such a relation permits a description of any correlation between *a* and *b*, particularly predicted by QM and observed in experiments. But in the frame of local realism the condition (1) is not indispensable. Instead, one can write

$$P_{ab}(\lambda_{1i},\lambda_{2i},\alpha,\beta) = P_a(\lambda_{1i},\alpha,\beta´) \times P_b(\lambda_{2i},\beta,\alpha´) \quad , \quad \text{(forecast)} \qquad (3)$$

where $\alpha´$ and $\beta´$ are the conditions of measurements in points ***A*** and ***B***, respectively, as they can be forecast by particles at the moment of their parting. If the forecast is good enough, i.e. $\alpha´ \approx \alpha$ and $\beta´ \approx \beta$, then (3) practically coincides with (2) and has all its advantages plus locality.

On the issue of separability: The EPR-particles have a common strategy. It can continue as long as they can forecast the future. But particles can also have so intensive interactions (e.g. with detectors) that initial strategy is not important anymore. In both cases the consciousness of the particle has an ability to cut off and forget the old partnership.

QM is "microsociology". Like its humane sister, it makes only probabilistic forecasts. The transition to classical physics is the transition from sociology of persons to sociology of crowds: the level of freedom decreases and behaviour becomes deterministic. Feynman's statement "quantum world is not like anything that we know" is right only if we do not take into account living beings. If a baby, having more experience with his parents than with "inanimate" matter, could make experiments, the behavior of microparticles would appear to it to be very natural.